\documentclass[a4paper,12pt]{article}
\usepackage{amsmath,amssymb,epsfig}

\makeatletter

\def\section{\@startsection{section}{1}{\z@}{-3.25ex plus -1ex minus
    -.2ex}{1.5ex plus .2ex}{\normalfont\bfseries}}

\renewenvironment{thebibliography}[1]
         {\section*{References}\frenchspacing\small
          \begin{list}{[\arabic{enumi}]}
         {\usecounter{enumi}\parsep=2pt\topsep 0pt
         \settowidth{\labelwidth}{[#1]}
         \leftmargin=\labelwidth\advance\leftmargin\labelsep
         \rightmargin=0pt\itemsep=0pt\sloppy}}{\end{list}}
\makeatother

\newcommand{\di}{\genfrac{}{}{0pt}{}}
\allowdisplaybreaks[4]

\newtheorem{thm}{Theorem}
\newtheorem{prp}[thm]{Proposition}

\begin{document}

\vspace*{-15mm}
\hspace*{\fill}hep-th/0403232

\vskip 10mm

\noindent
{\bfseries\large Renormalisation of $\phi^4$-theory on noncommutative 
$\mathbb{R}^4$ to all orders}
\\[2ex]
Harald {\sc Grosse}$^1$ and Raimar {\sc Wulkenhaar}$^2$
\\[0.5ex]
{\itshape\small 
$^{1}$\,Institut f\"ur Theoretische Physik, Universit\"at Wien,
Boltzmanngasse 5, A-1090 Wien, Austria. e-mail: harald.grosse@univie.ac.at
\\
$^{2}$\,Max-Planck-Institut f\"ur Mathematik in den
  Naturwissenschaften, 
Inselstra\ss{}e 22-26, D-04103 Leipzig, Germany. e-mail:
raimar.wulkenhaar@mis.mpg.de} 

\vskip .5cm

\noindent
{\bf Abstract.} We present the main ideas and techniques of the proof
that the duality-covariant four-dimensional noncommutative
$\phi^4$-model is renormalisable to all orders. This includes the
reformulation as a dynamical matrix model, the solution of the free
theory by orthogonal polynomials as well as the renormalisation by
flow equations involving power-counting theorems for ribbon graphs
drawn on Riemann surfaces.
\\[1ex]
\\[1ex]

\section{Introduction}

In recent years there has been considerable interest in quantum field
theories on the Moyal plane characterised by the $\star$-product 
(in $D$ dimensions)
\begin{align}
(a\star b)(x):= \int d^Dy \frac{d^Dk}{(2\pi)^D}
a(x{+}\frac{1}{2}\theta {\cdot} k) b(x{+}y)
\,\mathrm{e}^{\mathrm{i}ky}\;, \qquad
\theta_{\mu\nu}=-\theta_{\nu\mu} \in \mathbb{R}\;.
\end{align}
The interest was to a large extent
motivated by the observation that this kind of field theories arise
in the zero-slope limit of open string theory in presence of a
magnetic background field \cite{Seiberg:1999vs}.  A few months later
it was discovered \cite{Minwalla:1999px} (first for scalar models)
that these noncommutative field theories are not renormalisable beyond
a certain loop order.  The argument is that non-planar graphs are
finite but their amplitude grows beyond any bound when the external
momenta become exceptional. When inserted as subgraphs into bigger
graphs, these exceptional momenta are attained in the loop integration
and result in divergences for any number of external legs. This
problem is called UV/IR-mixing. A more rigorous explanation was given
in \cite{Chepelev:2000hm} where the problem was traced back to
divergences in some of the Hepp sectors which correspond to
\emph{disconnected} ribbon subgraphs wrapping the same handle of a
Riemann surface. Hepp sectors which correspond to connected non-planar
subgraphs are always finite.

The UV/IR-problem was found in all UV-divergent field theories on the
Moyal plane. Models with at most logarithmic UV-divergences (such as
two-dimensional and supersymmetric theories) can be defined at any
loop order, but their amplitudes are still unbounded at exceptional
momenta.

The UV/IR-mixing contains a clear
message: If we make the world noncommutative at very short
distances, we must---whether we like it or not---at the same time
modify the physics at large distances. The required modification is,
to the best of our knowledge, unique: It is given by an harmonic
oscillator potential for the free field action. In fact, we can prove
the following
\begin{thm}
The quantum field theory associated with the action 
\begin{align}
S= \int d^4x \Big( \dfrac{1}{2} \partial_\mu \phi \star \partial^\mu
\phi + \frac{\Omega^2}{2} (\tilde{x}_\mu \phi)
\star (\tilde{x}^\mu \phi) + \frac{\mu_0^2}{2} \phi \star \phi 
+ \frac{\lambda}{4!} \phi \star
\phi \star \phi \star \phi\Big)(x)\;,
\label{action}
\end{align}
for $\tilde{x}_\mu :=2(\theta^{-1})_{\mu\nu}\, x^\nu$, $\phi$-real,
Euclidean metric, is
perturbatively renormalisable to all orders in $\lambda$. 
\end{thm}
Our proof given in \cite{Grosse:2004yu} and \cite{Grosse:2003aj} is
very long so that there is some need of an introductory presentation
of its main ideas and techniques. 

First, we remark that the action is covariant with respect to a
\emph{duality} between position space and momentum 
space \cite{Langmann:2002cc}: Under the exchange of position
and momentum (i.e.\ not the Fourier transformation),
\begin{align}
p_\mu \leftrightarrow \tilde{x}_\mu\;,
\qquad \hat{\phi}(p) \leftrightarrow 
\pi^2 \sqrt{|\det \theta|}\;\phi(x)\;,
\end{align}
together with 
$\hat{\phi}(p_a)=\int d^4 x \;\mathrm{e}^{(-1)^a \mathrm{i}
  p_{a,\mu} x_a^\mu} \phi(x_a)$ for $a$ being a cyclic label, one has
\begin{align}
S\big[\phi;\mu_0,\lambda,\Omega\big] \mapsto 
\Omega^2
S\big[\phi;\frac{\mu_0}{\Omega},\frac{\lambda}{\Omega^2},\frac{1}{\Omega}
\big]\;.  
\label{Sdual}
\end{align}

\section{Reformulation as a dynamical matrix model}
\label{reformulation}

It is clear from the explicit $x$-dependence that for quantisation we
cannot proceed in momentum space. Fortunately, the Moyal plane has a
very convenient matrix base. 

We choose a coordinate frame where 
$\theta{=}\theta_{12}{=}{-}\theta_{21}{=}\theta_{34}{=}{-}\theta_{43}$ 
are the only non-vanishing $\theta$-components. We expanding the
fields according to 
$\phi(x) = \sum_{m^1,m^2,n^1,n^2 \in \mathbb{N}} 
\phi_{\di{m^1}{m^2}\di{n^1}{n^2}}
b_{\di{m^1}{m^2}\di{n^1}{n^2}}(x)$ where 
$b_{\di{m^1}{m^2}\di{n^1}{n^2}}(x)=
 f_{m^1n^1}(x_1,x_2) f_{m^2n^2}(x_3,x_4)$ with
\begin{align}
f_{m^1n^1}(x_1,x_2) &= 
\frac{(x_1{+}\mathrm{i}x_2)^{\star m^1}}{\sqrt{m^1!(2\theta)^{m^1}}} 
\star \Big(2 \mathrm{e}^{-\frac{1}{\theta}(x_1^2+x_2^2)}\Big) 
\star \frac{(x_1{-}\mathrm{i}x_2)^{\star n^1}}{\sqrt{n^1!(2\theta)^{n^1}}}
\;,
\\
(b_{mn} \star b_{kl})(x) &= \delta_{nk} b_{ml}(x)\;,\qquad 
\int d^4 x\, b_{mn}(x)= (2\pi \theta)^2\,\delta_{mn}\;.
\label{bb}
\end{align}
Due to (\ref{bb}) the non-local $\star$-product interaction becomes a
simple matrix product, at the price of rather complicated kinetic
terms and propagators. We obtain for the action (\ref{action})
\begin{align}
S & =(2\pi\theta)^2 
\sum_{m,n,k,l\in \mathbb{N}^2} \Big(\dfrac{1}{2} 
\phi_{mn} G_{mn;kl} \phi_{kl} + \frac{\lambda}{4!} 
\phi_{mn}\phi_{nk} \phi_{kl} \phi_{lm}\Big)\;,
\label{Sm}
\\
G_{\di{m^1}{m^2}\di{n^1}{n^2};\di{k^1}{k^2}\di{l^1}{l^2}} 
&= \big(\mu_0^2{+} \tfrac{2{+}2\Omega^2}{\theta}
(m^1{+}n^1{+}m^2{+}n^2{+}2) \big) \delta_{n^1k^1} \delta_{m^1l^1} 
\delta_{n^2k^2} \delta_{m^2l^2} 
\nonumber
\\
&- \tfrac{2{-}2\Omega^2}{\theta} \big(\sqrt{k^1l^1}\,
  \delta_{n^1+1,k^1}\delta_{m^1+1,l^1} + \sqrt{m^1n^1}\,
  \delta_{n^1-1,k^1} \delta_{m^1-1,l^1}\big)\delta_{n^2k^2}
  \delta_{m^2l^2} 
\nonumber
\\
&- \tfrac{2{-}2\Omega^2}{\theta} \big(\sqrt{k^2l^2}\,
  \delta_{n^2+1,k^2}\delta_{m^2+1,l^2} + \sqrt{m^2n^2}\,
  \delta_{n^2-1,k^2} \delta_{m^2-1,l^2}\big)\delta_{n^1k^1}
  \delta_{m^1l^1} \;.
\label{Gm}
\end{align}
Since the action is a trace we have $G_{mn;kl}=0$ unless $m{+}k=n{+}l$.

We are interested in a perturbative solution of the quantum field
theory around the free theory, the solution of which is given by the
propagator $\Delta_{mn;kl}$, i.e.\ the inverse of $G_{mn;kl}$. In a
first step we diagonalise the kinetic matrix: 
\begin{align}
G_{\di{m^1}{m^2}\di{m^1+\alpha^1}{m^2+\alpha^2};
\di{l^1+\alpha^1}{l^2+\alpha^2}\di{l^1}{l^2}}
&= \sum_{y^1,y^2=0}^\infty
U^{(\alpha^1)}_{m^1 y^1} U^{(\alpha^2)}_{m^2 y^2} 
 \big(\mu_0^2{+} \tfrac{4\Omega}{\theta}
(2y^1{+}2y^2{+}\alpha^1{+}\alpha^2{+}2) \big)
U^{(\alpha^1)}_{y^1 l^1} U^{(\alpha^2)}_{y^2 l^2} \;,
\label{G-diag}
\\
U^{(\alpha)}_{ny} 
&= \sqrt{\binom{\alpha{+}n}{n} \binom{\alpha{+}y}{y}}\; 
\Big(\frac{1{-}\Omega}{1{+}\Omega}\Big)^{n+y} 
\Big(\frac{2\sqrt{\Omega}}{1{+}\Omega}\Big)^{\alpha+1}\;
{}_2F_1\Big(\di{-n,-y}{1{+}\alpha}\big|\frac{4 \Omega}{
(1+\Omega)^2}\Big)\;.
\nonumber
\end{align}
For fixed $\alpha$, the kinetic matrix is in both components a Jacobi
matrix (a certain tridiagonal band matrix). The diagonalisation of
that band matrix yields the recursion relation for (orthogonal)
Meixner polynomials
$M_n(y;\beta,c)={}_2F_1\Big(\di{-n,-y}{\beta}\big|1{-}c\Big)$. The
corresponding equidistant eigenvalues are those of the harmonic
oscillator. To compute the propagator we have to invert the
eigenvalues $\big(\mu_0^2{+} \tfrac{4\Omega}{\theta}
(2y^1{+}2y^2{+}\alpha^1{+}\alpha^2{+}2) \big)$ in
(\ref{G-diag}). Using the identity
\begin{align}
\sum_{y=0}^\infty \frac{(\alpha{+}y)!}{y!\alpha!}\,  a^y\; &
  {}_2F_1\Big(\di{-m,-y}{1{+}\alpha}\Big| b\Big)
  \,{}_2F_1\Big(\di{-l,-y}{1{+}\alpha}\Big| b\Big) 
\nonumber
\\
&= \frac{(1{-}(1{-}b)a)^{m+l}}{(1{-}a)^{\alpha+m+l+1}}\;
  {}_2F_1\Big(\di{-m\,,\;-l}{1{+}\alpha}\Big| \frac{a
    b^2}{(1{-}(1{-}b)a)^2}\Big)\;,\qquad |a|<1\;,
\end{align}
which can be regarded as the heart of the renormalisation proof, we
arrive at 
\begin{align}
\Delta_{\di{m^1}{m^2}\di{n^1}{n^2};  
\di{k^1}{k^2}\di{l^1}{l^2}}
&= \frac{\theta}{2(1{+}\Omega)^2} \!
\sum_{v^1=\frac{|m^1-l^1|}{2}}^{\frac{m^1+l^1}{2}} 
\sum_{v^2=\frac{|m^2-l^2|}{2}}^{\frac{m^2+l^2}{2}} 
\!\! B\big(1{+} \tfrac{\mu_0^2 \theta}{8\Omega}
 {+}\tfrac{1}{2}(m^1{+}k^1{+}m^2{+}k^2){-}v^1{-}v^2,
1{+}2v^1{+}2v^2 \big)
\nonumber
 \\
&\times 
{}_2F_1\bigg(\di{1{+} 2v^1{+}2v^2\,,\; 
\frac{\mu_0^2 \theta}{8\Omega}
 {-}\frac{1}{2}(m^1{+}k^1{+}m^2{+}k^2){+}v^1{+}v^2
}{2{+} \frac{\mu_0^2 \theta}{8\Omega}
 {+}\frac{1}{2}(m^1{+}k^1{+}m^2{+}k^2){+}v^1{+}v^2} \bigg| 
\frac{(1{-} \Omega)^2}{(1{+}\Omega)^2}\bigg)
\Big(\frac{1{-} \Omega}{1{+}\Omega}\Big)^{2v^1+2v^2}
\nonumber
 \\
&\times 
\prod_{i=1}^2 \delta_{m^i+k^i,n^i+l^i}
\sqrt{ 
\binom{n^i}{v^i{+}\frac{n^i-k^i}{2}}
\binom{k^i}{v^i{+}\frac{k^i-n^i}{2}}
\binom{m^i}{v^i{+}\frac{m^i-l^i}{2}}
\binom{l^i}{v^i{+}\frac{l^i-m^i}{2}}}\;.
\label{prop}
\end{align}
One should appreciate here that the sum in (\ref{prop}) is finite,
i.e.\ we succeeded to solve the free theory with respect to the
preferred base of the interaction. The explicit solution enables a
fast numerical evaluation of the propagator, which is necessary to
determine the asymptotic behaviour of the propagator for large
indices. In few cases we can evaluate the sum exactly:
\begin{itemize}
\item $0\leq \Delta_{\di{m^1}{m^2}\di{n^1}{n^2};  
\di{k^1}{k^2}\di{l^1}{l^2}}(\mu_0) < \Delta_{\di{m^1}{m^2}\di{n^1}{n^2};  
\di{k^1}{k^2}\di{l^1}{l^2}}(0) $ 

This means that we can ignore the mass $\mu_0$
in our estimations for $\Omega>0$.

\item $\Delta_{\di{m}{0}\di{m}{0};  
\di{m}{0}\di{m}{0}}(0)
= \frac{\theta}{2(1{+}\Omega)^2(m{+}1)} \,{}_2F_1\Big(
\di{1,-m}{m{+}2}\Big|
\frac{(1{-} \Omega)^2}{(1{+}\Omega)^2}\Big) \sim 
\frac{\theta/8}{\Omega(m{+}1) + \sqrt{\frac{4}{\pi}(m{+}1)}}$

There is a discontinuity in the asymptotic behaviour of the propagator
at $\Omega=0$. For $\Omega=0$ there is a long-range correlation which
decays only very slowly with $\frac{1}{\sqrt{m}}$. This is the origin
of the UV/IR-mixing. For $\Omega>0$ the correlation decays with
$\frac{1}{m}$ which guarantees a good power-counting behaviour of the
model with $\Omega>0$. The asymptotic behaviour provides the easy part
of the renormalisation proof.

\item $\Delta_{\di{m^1}{m^2}\di{m^1}{m^2};\di{0}{0}\di{0}{0}}(0)
= \frac{\theta}{2(1{+}\Omega)^2(m^1{+}m^2{+}1)} 
\big(\frac{1{-} \Omega}{1{+}\Omega}\big)^{m^1+m^2}$ 

This property controls the non-locality. The model is non-local in the
sense that there is a correlation $\Delta_{mn;kl}$ for arbitrarily large
$\|m-l\|$. However, that correlation is exponentially suppressed,
preserving some sort of quasi-locality. This provides the tricky part
of the  renormalisation proof.

\end{itemize}

\section{The Polchinski equation}
\label{Polchinski}

It is, in principle, possible to proceed with the discussion of
Feynman graphs built with the propagator (\ref{prop}) according to
Zimmermann's forest formula. But the complexity of the arising graphs
(compare (\ref{prop}) with the simple $\frac{1}{k^2+m^2}$ of
commutative field theories) requires a more sophisticated approach:
the renormalisation by flow equations. The idea goes back to Wilson
\cite{Wilson:1973jj} and was further developed by Polchinski to an
efficient renormalisation proof of commutative $\phi^4$-theory
\cite{Polchinski:1983gv}.

The starting point is the definition of the quantum field theory by
the cut-off partition function
\begin{align}
Z[J,\Lambda]&= \displaystyle \int \Big(\prod_{a,b}
  d \phi_{ab}\Big) \,\exp\big(-S[\phi,J,\Lambda]\big)\;,
\label{ZJL}
  \\*
S[\phi,J,\Lambda] &= (2\pi\theta)^2 \Big( \sum_{m,n,k,l} \frac{1}{2}
  \phi_{mn} G^K_{mn;kl} (\Lambda)\, \phi_{kl} + L[\phi,\Lambda] +
  C[\Lambda]
\nonumber
  \\*[-2ex]
  & \hspace*{10em} +\sum_{m,n,k,l} \phi_{mn} F_{mn;kl}[\Lambda] J_{kl} +
  \sum_{m,n,k,l} \frac{1}{2} J_{mn} E_{mn;kl}[\Lambda] J_{kl} \Big)\;.
\label{Z}
\end{align}
The most important pieces here are the cut-off kinetic term
\begin{align}
G^K_{\di{m^1}{m^2}\di{n^1}{n^2};\di{k^1}{k^2}\di{l^1}{l^2}}(\Lambda) 
&:= \hspace*{-1em}  
\prod_{\mbox{\tiny$\begin{array}{ll}
i \in \hspace*{-1.5\tabcolsep} & m^1,m^2,n^1,n^2,\\ & 
k^1,k^2,l^1,l^2\end{array}$}} \hspace*{-1em}  
K^{-1}\big(\tfrac{ i }{\theta \Lambda^2}\big) \; 
G_{\di{m^1}{m^2}\di{n^1}{n^2};\di{k^1}{k^2}\di{l^1}{l^2}}\;,
\parbox{60mm}{\begin{picture}(40,30)
 \put(-3,-97.5){\epsfig{file=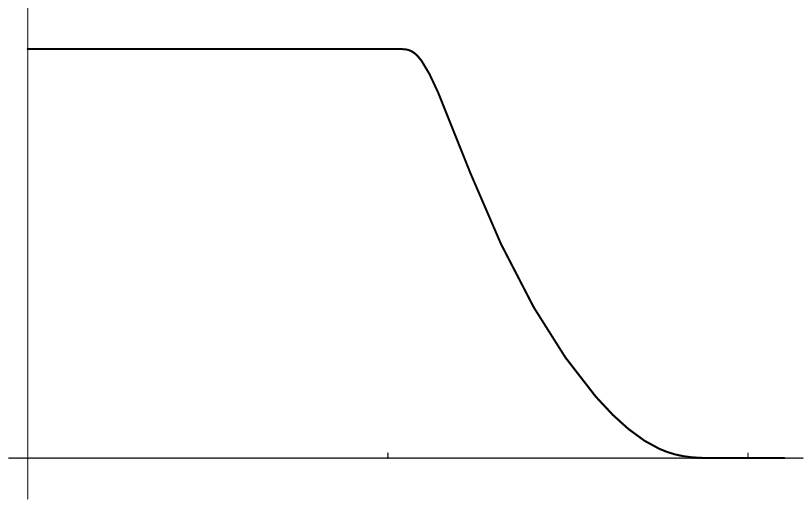,scale=0.45,bb=0 0 598 843}}
 \put(7,27.5){\mbox{\scriptsize$K[\frac{i}{\theta\Lambda^2}]$}}
 \put(8,3){\mbox{\scriptsize$0$}}
 \put(8,22){\mbox{\scriptsize$1$}}
 \put(51,4){\mbox{\scriptsize$i$}}
 \put(26,0){\mbox{\scriptsize$\theta \Lambda^2$}}
 \put(42,0){\mbox{\scriptsize$2 \theta \Lambda^2$}}
 \end{picture}}
\end{align}
where the weight of the matrix indices is altered according to a
\emph{smooth} cut-off function\footnote{We understand the cut-off as a
  limiting process $\epsilon \to 0$ in
  $K^{-1}(\frac{i}{\theta\Lambda^2})=\frac{1}{\epsilon}$ for $i\geq
  2\theta\Lambda^2$. In the limit, the partition function (\ref{ZJL})
  vanishes unless $\phi_{\di{m^1}{m^2}\di{n^1}{n^2}}=0$ if
  $\max(m^1,m^2,n^1,n^2)\geq 2\theta\Lambda^2$, thus implementing a
  cut-off of the measure $\prod_{a,b} d\phi_{ab}$ in (\ref{ZJL}). All
  other formulae involve $K(\frac{i}{\theta\Lambda^2})$.} $K$, and the
\emph{effective action} $L[\phi,\Lambda]$ which compensates the effect
of the cut-off. We are interested in the limit $\Lambda \to \infty$,
where the cut-off goes away, $\lim_{\Lambda \to \infty}
K[\frac{i}{\theta\Lambda^2}]=1$.  Thus, we would formally obtain the
original model for $\Lambda=\infty$ and $L[\phi,\infty] =
\frac{\lambda}{4!}\sum_{m,n,k,l}
\phi_{mn}\phi_{nk}\phi_{kl}\phi_{lm}$, $C[\infty]=0$,
$E_{mn;kl}[\infty]=0$, $F_{mn;kl}[\infty]=\delta_{nk} \delta_{ml}$.
However, $\Lambda=\infty$ is difficult to obtain due to the appearance
of divergences, which require compensating counterterms in $L[\phi]$.

The genial idea of the renormalisation group approach is to require
instead the \emph{independence} of the partition function on the
cut-off, $\Lambda \frac{\partial}{\partial \Lambda} Z
[J,\Lambda]=0$. Working out the details one arrives, in particular, at
the Polchinski equation for matrix models
\begin{align}
\Lambda \frac{\partial L[\phi,\Lambda]}{\partial  \Lambda} 
&= \sum_{m,n,k,l} \frac{1}{2} \Lambda \frac{\partial
  \Delta^K_{nm;lk}(\Lambda)}{\partial \Lambda} \Big( \frac{\partial
  L[\phi,\Lambda]}{\partial \phi_{mn}}\frac{\partial
  L[\phi,\Lambda]}{\partial \phi_{kl}} - \frac{1}{(2\pi \theta)^2}
\frac{\partial^2 L[\phi,\Lambda]}{\partial \phi_{mn}\,\partial
  \phi_{kl}}\Big)\;,
\label{polch}
\end{align}
where $\Delta^K_{nm;lk}(\Lambda) := \prod_{i\in m^1,m^2,\dots,l^1,l^2}
K(\frac{i }{\theta \Lambda^2}) \Delta_{nm;lk}$. To obtain
(\ref{polch}) it was, of course,
important to realise finite matrices via a smooth function $K$. There
are other differential equations for the functions $C,E,F$ in
(\ref{Z}) which, however, are trivial to integrate. The true
difficulties are contained in the non-linear differential equation
(\ref{polch}). 

The Polchinski equation has a non-perturbative meaning, but to solve
it we need, for the time being, a power series ansatz: 
\begin{align}
L[\phi,\Lambda]
= \sum_{V=1}^\infty \lambda^V
\sum_{N=2}^{2V+2} \frac{(2\pi\theta)^{\frac{N}{2}-2}}{N!} 
\sum_{m_1,n_i \in \mathbb{N}^2} A^{(V)}_{m_1n_1;\dots;m_Nn_N}[\Lambda]
\phi_{m_1n_1} \cdots  \phi_{m_Nn_N} \;.
\end{align}
Then, the differential equation (\ref{polch}) provides an explicit
recursive solution for the coefficients
$A^{(V)}_{m_1n_1;\dots;m_Nn_N}[\Lambda]$ which, because the fields 
$\phi_{mn}$ carry two indices, is represented by \emph{ribbon
  graphs}: 
\begin{align}
\Lambda \frac{\partial}{\partial \Lambda}
&\parbox{26mm}{\begin{picture}(20,21) 
 \put(0,0){\epsfig{file=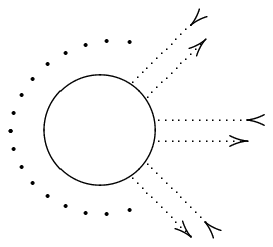,bb=69 621 141 684}}
 \put(22,13.5){\mbox{\scriptsize$n_1$}}
 \put(20,8){\mbox{\scriptsize$m_1$}}
 \put(21,3){\mbox{\scriptsize$n_2$}}
 \put(13,1){\mbox{\scriptsize$m_2$}}
 \put(19,17){\mbox{\scriptsize$m_N$}}
 \put(15,23){\mbox{\scriptsize$n_N$}}
\end{picture}}
\nonumber
\\*[-4ex]
&
=\frac{1}{2}\sum_{m,n,k,l} \sum_{N_1=1}^{N-1}
\parbox{48mm}{\begin{picture}(48,25) 
 \put(0,0){\epsfig{file=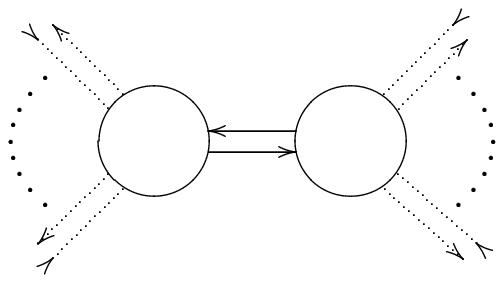,bb=69 615 214 684}}
 \put(0,4){\mbox{\scriptsize$m_1$}}
 \put(6,0){\mbox{\scriptsize$n_1$}}
 \put(0,21){\mbox{\scriptsize$n_{N_1}$}}
 \put(7,23){\mbox{\scriptsize$m_{N_1}$}}
 \put(47,20){\mbox{\scriptsize$m_{N_1+1}$}}
 \put(37,26){\mbox{\scriptsize$n_{N_1+1}$}}
 \put(48,3){\mbox{\scriptsize$n_N$}}
 \put(39,2){\mbox{\scriptsize$m_N$}}
 \put(27,14.5){\mbox{\scriptsize$k$}}
 \put(28,8){\mbox{\scriptsize$l$}}
 \put(22,15){\mbox{\scriptsize$n$}}
 \put(22,9){\mbox{\scriptsize$m$}}
\end{picture}}
\quad -  \frac{1}{4\pi \theta}
\sum_{m,n,k,l} \parbox{35mm}{\begin{picture}(30,35) 
 \put(0,0){\epsfig{file=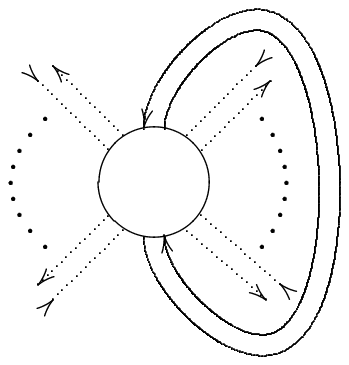,bb=69 585 168 687}}
 \put(0,9){\mbox{\scriptsize$m_1$}}
 \put(6,5){\mbox{\scriptsize$n_1$}}
 \put(-1,26){\mbox{\scriptsize$n_{i-1}$}}
 \put(7.5,29){\mbox{\scriptsize$m_{i-1}$}}
 \put(27,25){\mbox{\scriptsize$m_i$}}
 \put(22.5,30){\mbox{\scriptsize$n_i$}}
 \put(27,9){\mbox{\scriptsize$n_N$}}
 \put(20,7){\mbox{\scriptsize$m_N$}}
 \put(12,25.5){\mbox{\scriptsize$n$}}
 \put(18,26){\mbox{\scriptsize$m$}}
 \put(12,10.5){\mbox{\scriptsize$k$}}
 \put(18,10.5){\mbox{\scriptsize$l$}}
\end{picture}}
\label{polLgraph}
\end{align}
An internal double line symbolises the propagator $Q_{mn;kl}(\Lambda)
:= \frac{1}{2\pi\theta} 
\Lambda \frac{\partial}{\partial
\Lambda} \Delta^K_{mn;kl}(\Lambda)
= \parbox{12mm}{\begin{picture}(10,8)
   \put(0,3){\epsfig{file=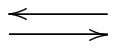,bb=71 667 101 675}}
 \put(2,6.5){\mbox{\scriptsize$n$}}
 \put(6,1){\mbox{\scriptsize$l$}}
 \put(0,1){\mbox{\scriptsize$m$}}
 \put(8,6.5){\mbox{\scriptsize$k$}}
\end{picture}}$.
   
Clearly, in this way we produce very complicated ribbon graphs which
cannot be drawn any more in a plane. Ribbon graphs define a Riemann
surface on which they can be drawn. The Riemann surface is
characterised by its genus $g$ computable via the Euler characteristic
of the graph, $g=1-\frac{1}{2}(L-I+V)$, and the number $B$ of
holes. Here, $L$ is the number of single-line loops if we close the
external lines of the graph, $I$ is the number of double-line
propagators and $V$ the number of vertices. The number $B$ of holes
coincides with the number of single-line cycles which carry external
legs. A few examples might help to understand the closure of external
lines and the resulting topological data:
\begin{align}
\parbox{45mm}{\begin{picture}(40,35) 
      \put(0,0){\epsfig{file=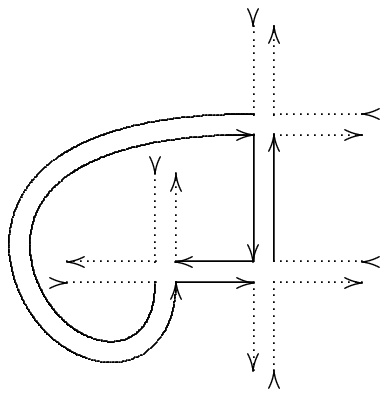,bb=74 583 178 684}}
      \put(7,7.5){\mbox{\scriptsize$n_1$}}
      \put(9,12.5){\mbox{\scriptsize$m_1$}}
      \put(30,7){\mbox{\scriptsize$m_3$}}
      \put(32,12.5){\mbox{\scriptsize$n_3$}}
      \put(30,22){\mbox{\scriptsize$m_4$}}
      \put(32,27.5){\mbox{\scriptsize$n_4$}}
      \put(21,4){\mbox{\scriptsize$m_2$}}
      \put(28,2){\mbox{\scriptsize$n_2$}}
      \put(21.5,34){\mbox{\scriptsize$n_5$}}
      \put(28,32){\mbox{\scriptsize$m_5$}}
      \put(11.5,19){\mbox{\scriptsize$n_6$}}
      \put(18,17){\mbox{\scriptsize$m_6$}}
    \end{picture}}
  \quad \Rightarrow \quad 
\parbox{35mm}{\begin{picture}(30,30) 
  \put(0,0){\epsfig{file=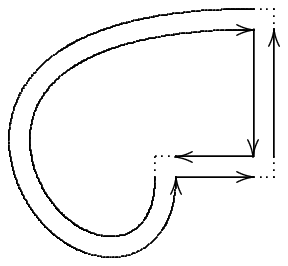,bb=74 610 153 684}}
\end{picture}}
\qquad 
\begin{array}{r@{\,}l@{\qquad}r@{\,}ll}
L &= 2  & g &= 0  \\
I &= 3  & B &= 2 \\
V &= 3  & N &= 6 
\end{array}
\label{gt1}
\\[1ex]
\parbox{50mm}{\begin{picture}(30,25)
    \put(0,0){\epsfig{file=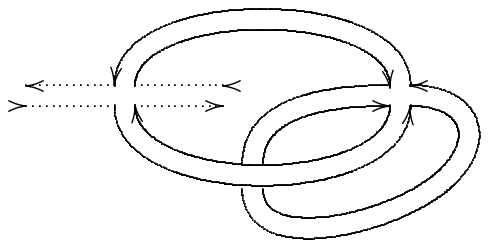,bb=71 612 204 681}}
    \put(0,11.5){\mbox{\scriptsize$n_1$}}
    \put(3,17.5){\mbox{\scriptsize$m_1$}}
    \put(15,11.5){\mbox{\scriptsize$m_2$}}
    \put(17,17.5){\mbox{\scriptsize$n_2$}}
\end{picture}}
 \quad \Rightarrow \quad 
\parbox{35mm}{\begin{picture}(30,25) 
  \put(0,0){\epsfig{file=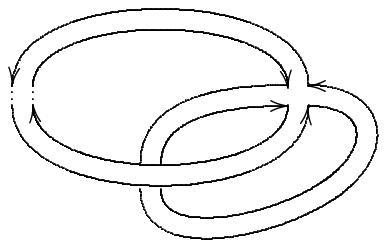,bb=71 612 178 681}}
\end{picture}}
\qquad 
\begin{array}{r@{\,}l@{\qquad}r@{\,}ll}
L &= 1  & g &= 1 \\
I &= 3  & B &= 1  \\
V &= 2  & N &= 2 
\end{array}
\label{gt3}
\end{align}
According to the topology we label the expansion coefficients of the
effective action by $A^{(V,B,g)}_{m_1n_1;\dots;m_Nn_N}$.

\section{Integration procedure of the Polchinski equation}
\label{integration} 

The integration procedure of the Polchinski equation involves the
entire magic of renormalisation. Suppose we want to evaluate
the planar one-particle irreducible four-point function with two
vertices, $A^{(2,1,0)\text{1PI}}_{m_1n_1;\dots;m_Nn_N}$. The
Polchinski equation (\ref{polLgraph}) provides the
$\Lambda$-derivative of that function:
\begin{align}
\Lambda \frac{\partial}{\partial \Lambda} 
A^{(2,1,0)\text{1PI}}_{mn;nk;kl;lm}[\Lambda] = \sum_{p \in
  \mathbb{N}^2} \left(\quad
  \parbox{38mm}{\begin{picture}(38,15)
       \put(0,0){\epsfig{file=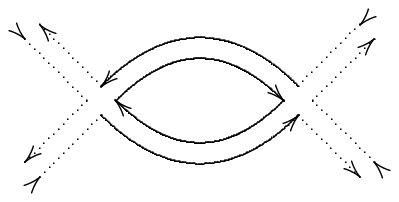,bb=71 630 174 676}}
       \put(-4,12){\mbox{\footnotesize$m$}}
       \put(-1,6){\mbox{\footnotesize$m$}}
       \put(34,9){\mbox{\footnotesize$k$}}
       \put(36,4){\mbox{\footnotesize$k$}}
       \put(4,-1){\mbox{\footnotesize$n$}}
       \put(30,-1){\mbox{\footnotesize$n$}}
       \put(5,14){\mbox{\footnotesize$l$}}
       \put(30,15){\mbox{\footnotesize$l$}}
       \put(12,8){\mbox{\footnotesize$p$}}
       \put(22,8){\mbox{\footnotesize$p$}}
   \end{picture}} \right)(\Lambda) + \text{ permutations}\;.
\label{A4diff}
\end{align}
We consider the special case with constant indices on the
trajectories. The first guess would be to perform the
$\Lambda$-integration of (\ref{A4diff}) from some initial scale
$\Lambda_0$ (sent to $\infty$ at the end) down to $\Lambda$. However,
this choice of integration leads to
$A^{(2,1,0)\text{1PI}}_{mn;nk;kl;lm}[\Lambda] \sim \ln
\frac{\Lambda_0}{\Lambda}$, which diverges when we remove the cutoff
$\Lambda_0\to\infty$. Following Polchinski we understand
renormalisation as the change of the boundary condition for the
integration. Thus, the idea would be to introduce a renormalisation
scale $\Lambda_R$ so that we would integrate (\ref{A4diff}) from
$\Lambda_R$ up to $\Lambda$. Then,
$A^{(2,1,0)\text{1PI}}_{mn;nk;kl;lm}[\Lambda] \sim \ln
\frac{\Lambda}{\Lambda_R}$, and there would be no problem any more sending
$\Lambda_0\to\infty$. However, since there is an \emph{infinite number} of
matrix indices and there is no symmetry which could relate the
amplitudes for different indices, that integration procedure entails an
infinite number of initial conditions
$A^{(2,1,0)\text{1PI}}_{mn;nk;kl;lm}[\Lambda_R]$. These initial
conditions correspond to normalisation experiments, and clearly a
model requiring an infinite number of normalisation experiments has no
physical meaning. Thus, to have a renormalisable model, we can only
afford a finite number of integrations from $\Lambda_R$ up to
$\Lambda$. The discussion shows that the correct integration procedure
is something like
\begin{align}
&A^{(2,1,0)\text{1PI}}_{mn;nk;kl;lm}[\Lambda]
\nonumber
\\*[-1ex]
&\quad =  -\int_{\Lambda}^{\Lambda_0} 
\frac{d \Lambda'}{\Lambda'} \,
\sum_{p \in \mathbb{N}^2} \left(~~~
\parbox{39mm}{\begin{picture}(20,15)
       \put(0,0){\epsfig{file=a24,bb=71 630 174 676}}
       \put(-4,12){\mbox{\footnotesize$m$}}
       \put(-1,6){\mbox{\footnotesize$m$}}
       \put(34,9){\mbox{\footnotesize$k$}}
       \put(36,4){\mbox{\footnotesize$k$}}
       \put(4,-1){\mbox{\footnotesize$n$}}
       \put(30,-1){\mbox{\footnotesize$n$}}
       \put(5,14){\mbox{\footnotesize$l$}}
       \put(30,15){\mbox{\footnotesize$l$}}
       \put(12,8){\mbox{\footnotesize$p$}}
       \put(22,8){\mbox{\footnotesize$p$}}
   \end{picture}}
-~~
\parbox{40mm}{\begin{picture}(20,15)
       \put(0,0){\epsfig{file=a24,bb=71 630 174 676}}
       \put(-4,12){\mbox{\footnotesize$m$}}
       \put(-1,6){\mbox{\footnotesize$m$}}
       \put(34,9){\mbox{\footnotesize$k$}}
       \put(36,4){\mbox{\footnotesize$k$}}
       \put(4,-1){\mbox{\footnotesize$n$}}
       \put(30,-1){\mbox{\footnotesize$n$}}
       \put(5,14){\mbox{\footnotesize$l$}}
       \put(30,15){\mbox{\footnotesize$l$}}
       \put(8,12){\mbox{\footnotesize$0$}}
       \put(25.5,12){\mbox{\footnotesize$0$}}
       \put(8,2){\mbox{\footnotesize$0$}}
       \put(25.5,2){\mbox{\footnotesize$0$}}
       \put(12,8){\mbox{\footnotesize$p$}}
       \put(22,8){\mbox{\footnotesize$p$}}
   \end{picture}}
\right)\![\Lambda']
\nonumber
\\*[2ex]
&\quad + ~~\parbox{20mm}{\begin{picture}(20,15)
       \put(0,0){\epsfig{file=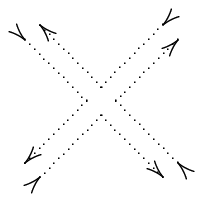,bb=71 638 117 684}}
   \put(-4,12){\mbox{\footnotesize$m$}}
       \put(-1,6){\mbox{\footnotesize$m$}}
       \put(13,9){\mbox{\footnotesize$k$}}
       \put(14,4){\mbox{\footnotesize$k$}}
       \put(4,-1){\mbox{\footnotesize$n$}}
       \put(10,-1){\mbox{\footnotesize$n$}}
       \put(5,14){\mbox{\footnotesize$l$}}
       \put(10,15){\mbox{\footnotesize$l$}}
   \end{picture}}  \left[
\int_{\Lambda_R}^\Lambda \frac{d \Lambda'}{\Lambda'} \, 
\sum_{p \in \mathbb{N}^2} 
\left(~~\parbox{40mm}{\begin{picture}(20,15)
       \put(0,0){\epsfig{file=a24,bb=71 630 174 676}}
       \put(-2,11){\mbox{\footnotesize$0$}}
       \put(-1,5){\mbox{\footnotesize$0$}}
       \put(34,9){\mbox{\footnotesize$0$}}
       \put(36,4){\mbox{\footnotesize$0$}}
       \put(4,-1){\mbox{\footnotesize$0$}}
       \put(29,-1){\mbox{\footnotesize$0$}}
       \put(5,14){\mbox{\footnotesize$0$}}
       \put(30,15){\mbox{\footnotesize$0$}}
       \put(12,8){\mbox{\footnotesize$p$}}
       \put(22,8){\mbox{\footnotesize$p$}}
   \end{picture}}\right)\![\Lambda']
+ A^{(2,1,0)\text{1PI}}_{00;00;00;00}[\Lambda_R] \right]\,.
\label{A4}
\end{align}
The second graph in the first line on the rhs and the graph in
brackets in the last line are identical, because only the indices on
the propagators determine the value of the graph. Moreover, the vertex
in the last line in front of the bracket equals $1$. Thus,
differentiating (\ref{A4}) with respect to $\Lambda$ we obtain indeed
(\ref{A4diff}). As a further check one can consider (\ref{A4}) for
$m=n=k=l=0$. Finally, the independence of
$A^{(2,1,0)\text{1PI}}_{mn;nk;kl;lm}[\Lambda_0]$ 
on the indices $m,n,k,l$ is built-in. This property is, for
$\Lambda_0 \to \infty$, dynamically generated by the model. 

There is a similar $\Lambda_0$-$\Lambda_R$-mixed integration procedure
for the planar 1PI two-point functions
$A^{(V,1,0)\text{1PI}}_{\di{m^1}{m^2}\di{n^1}{n^2};
  \di{n^1}{n^2}\di{m^1}{m^2}}$,
$A^{(V,1,0)\text{1PI}}_{\di{m^1+1}{m^2}\di{n^1+1}{n^2};
  \di{n^1}{n^2}\di{m^1}{m^2}}$ and
$A^{(V,1,0)\text{1PI}}_{\di{m^1}{m^2+1}\di{n^1}{n^2+1};
  \di{n^1}{n^2}\di{m^1}{m^2}}$. These involve in total three different
sub-integrations from $\Lambda_R$ up to $\Lambda$. We refer to
\cite{Grosse:2004yu} for details. All other graphs are integrated from
$\Lambda_0$ down to $\Lambda$, e.g.\ 
\begin{align}
A^{(2,2,0)\text{1PI}}_{m_1n_1;\dots;m_4n_4}[\Lambda]  
=    -\int_{\Lambda}^{\Lambda_0} 
\frac{d \Lambda'}{\Lambda'} \,
\sum_{p \in \mathbb{N}^2} \left( ~
\parbox{42mm}{\begin{picture}(20,20)
       \put(0,0){\epsfig{file=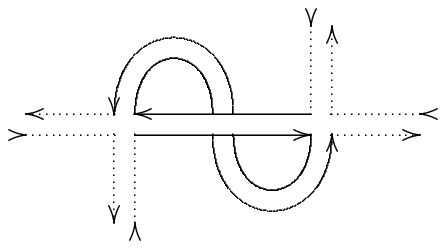,bb=71 625 187 684}}
       \put(2,13){\mbox{\footnotesize$m_4$}}
       \put(0,7){\mbox{\footnotesize$n_4$}}
       \put(4,2){\mbox{\footnotesize$m_1$}}
       \put(13,0){\mbox{\footnotesize$n_1$}}
       \put(36,13){\mbox{\footnotesize$n_2$}}
       \put(34,7){\mbox{\footnotesize$m_2$}}
       \put(32,18){\mbox{\footnotesize$m_3$}}
       \put(24,20){\mbox{\footnotesize$n_3$}}
       \put(13.5,13.5){\mbox{\footnotesize$p$}}
   \end{picture}} \right)\![\Lambda']\;.
\label{A220}
\end{align}

\section{The power-counting theorem}
\label{power-counting}

\begin{thm}
The previous integration procedure yields
\begin{align}
\big|A^{(V,B,g)}_{
    m_1n_1;\dots;m_Nn_N}[\Lambda] \big| 
\leq \big(\sqrt{\theta} \Lambda\big)^{(4-N)+4(1-B-2g)}\;
P^{2V-\frac{N}{2}}\Big[\ln \frac{\Lambda}{\Lambda_R}\Big]\;,
\label{pc}
\end{align}
where $P^q[X]$ stands for a polynomial of degree $q$ in $X$.
\end{thm}
\emph{Idea of the proof.} The cut-off propagator $Q_{mn;kl}(\Lambda)$
contains both an UV and an IR cut-off,
$Q_{\di{m_1}{m_2}\di{n_1}{n_2};\di{k_1}{k_2}\di{l_1}{l_2}}(\Lambda)
\neq 0$ only for $\theta \Lambda^2 < \max(m_1,\dots,l_2) < 2 \theta
\Lambda^2$. The global maximum of the propagator $\Delta_{mn;kl}$ is
at $m=n=k=l=\di{0}{0}$. If $\Lambda$ increases, at least one of the
indices of $Q_{mn;kl}$ must increase as well, resulting in a decrease
of $\big|Q_{mn;kl}(\Lambda)\big|$ with $\Lambda$. If we normalise the
volume of the support of $Q_{mn;kl}(\Lambda)$ with respect to a single
index to $\theta^2\Lambda^4$ (corresponding to a four-dimensional
model), then
\begin{align}
|Q_{mn;kl}(\Lambda)| < \frac{C_0}{\Omega \theta \Lambda^2} \delta_{m+k,n+l} \;.
\label{Qmax}
\end{align}
Thus, the propagator and the volume of a loop summation have the same
power-counting dimensions as a commutative $\phi^4$-model in momentum
space, giving the total power-counting degree $4-N$ for an $N$-point
function. 

This is (more or less, see below) correct for planar graphs. The
scaling behaviour of non-planar graphs is considerably improved by the
\emph{anisotropy} (or quasi-locality) of the propagator:
\begin{align}
\parbox{130mm}{\begin{picture}(120,40)
\put(-25,-105){\epsfig{file=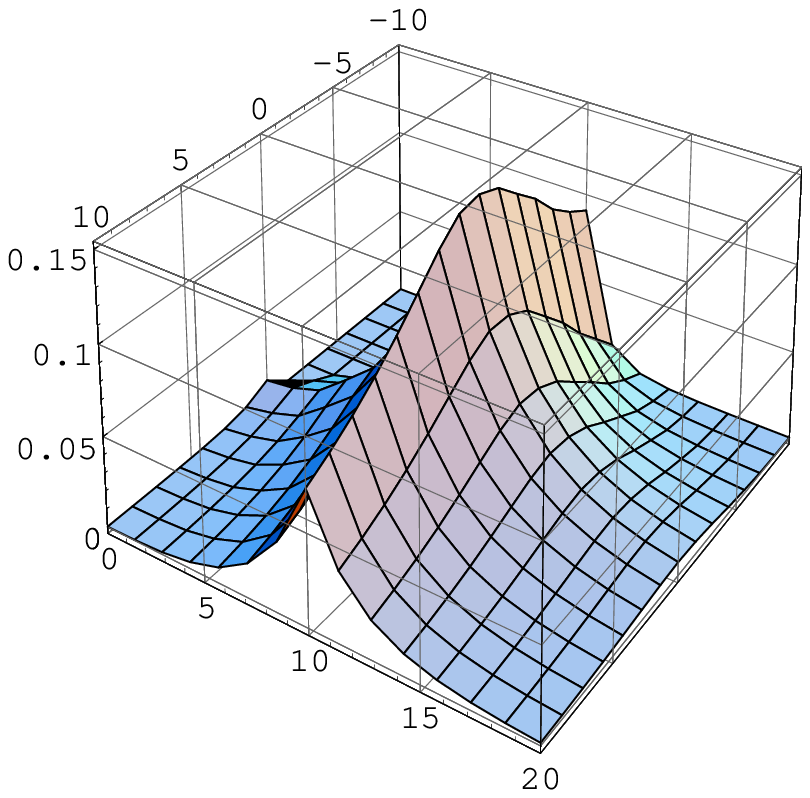,scale=0.55,bb=0 0 598 843}}
\put(80,10){\vector(0,1){20}}
\put(80,10){\vector(4,-1){15}}
\put(80,10){\vector(3,1){10}}
\put(64,33){\mbox{\footnotesize$\theta^{-1} 
\Delta_{\di{10}{0}\di{10+\alpha}{0};\di{l+\alpha}{0}
\di{l}{0}}$}}
\put(92,15){\mbox{\footnotesize$\alpha$}}
\put(97,5){\mbox{\footnotesize$l$}}
\put(40,0){\mbox{\footnotesize$\Omega=0.1$}}
\put(55,0){\mbox{\footnotesize$\mu_0=0$}}
\end{picture}}
\label{quasi-local}
\end{align}
As a consequence, for given index $m$ of the propagator 
$Q_{mn;kl}(\Lambda)
= \parbox{12mm}{\begin{picture}(10,8)
   \put(0,3){\epsfig{file=p1,bb=71 667 101 675}}
 \put(2,6.5){\mbox{\scriptsize$n$}}
 \put(6,1){\mbox{\scriptsize$l$}}
 \put(0,1){\mbox{\scriptsize$m$}}
 \put(8,6.5){\mbox{\scriptsize$k$}}
\end{picture}}$, the contribution to a graph is strongly suppressed
unless the other index $l$ on the trajectory through $m$ is close to
$m$. Thus, the sum over $l$ for given $m$ converges and does not alter
(apart from a factor $\Omega^{-1}$) the power-counting behaviour of
(\ref{Qmax}):
\begin{align}
\sum_{l \in \mathbb{N}^2} \Big(\max_{n,k}  |Q_{mn;kl}(\Lambda)| \Big) 
< \frac{C_1}{\theta\Omega^2 \Lambda^2}  \;.
\end{align}
In a non-planar graph like the one in (\ref{A220}), the index
$n_3$---fixed as an external index---localises the summation index
$p\approx n_3$. Thus, we save one volume factor $\theta^2\Lambda^4$ 
compared with a true loop summation as in (\ref{A4}). In general, each
hole in the Riemann surface saves one volume factor, and each handle
even saves two: In the genus-$1$ graph
\begin{align}
\sum_{p,q,r \in \mathbb{N}^2} \quad
\parbox{53mm}{\begin{picture}(30,27)
       \put(0,0){\epsfig{file=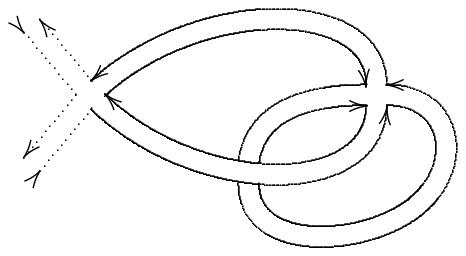,bb=71 608 198 679}}
       \put(4,21){\mbox{\footnotesize$m_1$}}
       \put(-4,20){\mbox{\footnotesize$n_1$}}
       \put(3,6){\mbox{\footnotesize$n_2$}}
       \put(-4,13){\mbox{\footnotesize$m_2$}}
       \put(31.5,18.5){\mbox{\footnotesize$r$}}
       \put(31.5,12){\mbox{\footnotesize$q$}}
       \put(38,12){\mbox{\footnotesize$p$}}
   \end{picture}}
\label{g1}
\end{align}
$n_2$ is fixed as an external index, and the quasi-locality
(\ref{quasi-local}) implies $n_2 \approx p \approx q \approx r$. Thus,
instead of the two loops of a corresponding line graph, the non-planar
ribbon graph (\ref{g1}) does not require any volume factor in the
power-counting estimation. 

A more careful analysis of (\ref{prop}) shows that also planar graphs
get suppressed  with $\big|Q_{\di{m^1}{m^2}\di{n^1}{n^2};  
\di{k^1}{k^2}\di{l^1}{l^2}}(\Lambda)\big|
< \frac{ C_2}{\Omega \theta \Lambda^2} 
\prod_{i=1}^2\big(\frac{\max(m^i,l^i)+1}{\theta
  \Lambda^2}\big)^{\frac{|m^i-l^i|}{2}}$, for $m^i\leq n^i$,
if the index along a trajectory jumps. This leaves the
functions $A^{(V,1,0)\text{1PI}}_{mn;nk;kl;lm}$,
$A^{(V,1,0)\text{1PI}}_{\di{m^1}{m^2}\di{n^1}{n^2};
\di{n^1}{n^2}\di{m^1}{m^2}}$,
$A^{(V,1,0)\text{1PI}}_{\di{m^1+1}{m^2}\di{n^1+1}{n^2};
\di{n^1}{n^2}\di{m^1}{m^2}}$ and 
$A^{(V,1,0)\text{1PI}}_{\di{m^1}{m^2+1}\di{n^1}{n^2+1};
\di{n^1}{n^2}\di{m^1}{m^2}}$ as the only relevant or marginal ones. In
these functions one has to use a discrete version of the Taylor
expansion, 
\begin{align}
\Big|Q_{\di{m^1}{m^2}\di{n^1}{n^2};  
\di{n^1}{n^2}\di{m^1}{m^2}}(\Lambda)
&- Q_{\di{0}{0}\di{n^1}{n^2};  
\di{n^1}{n^2}\di{0}{0}}(\Lambda) \Big| < 
\frac{C_3}{\Omega \theta \Lambda^2} 
\Big(\frac{\max(m^1,m^2)}{\theta \Lambda^2}\Big)\;,
\\
\Big|Q_{\di{m^1}{m^2}\di{n^1}{n^2};  
\di{n^1}{n^2}\di{m^1}{m^2}}(\Lambda)
&- Q_{\di{0}{0}\di{n^1}{n^2};  
\di{n^1}{n^2}\di{0}{0}}(\Lambda)
- m^1 \big(Q_{\di{1}{0}\di{n^1}{n^2};  
\di{n^1}{n^2}\di{1}{0}}(\Lambda)
-Q_{\di{0}{0}\di{n^1}{n^2};  
\di{n^1}{n^2}\di{0}{0}}(\Lambda)\big)
\nonumber
\\
&- m^2 \big(Q_{\di{0}{1}\di{n^1}{n^2};  
\di{n^1}{n^2}\di{0}{1}}(\Lambda)
-Q_{\di{0}{0}\di{n^1}{n^2};  
\di{n^1}{n^2}\di{0}{0}}(\Lambda)\big)\Big|
< \dfrac{C_4}{\Omega \theta \Lambda^2} 
\Big(\frac{\max(m^1,m^2)}{\theta \Lambda^2}\Big)^2\;,
\\
\Big|Q_{\di{m^1+1}{m^2}\di{n^1+1}{n^2};  
\di{n^1}{n^2}\di{m^1}{m^2}}(\Lambda)
&- \sqrt{m^1{+}1} Q_{\di{1}{0}\di{n^1+1}{n^2};  
\di{n^1}{n^2}\di{0}{0}}(\Lambda) \Big| 
< \frac{C_5}{\Omega \theta \Lambda^2} 
\Big(\frac{\max(m^1,m^2)}{\theta \Lambda^2}\Big)^{\frac{3}{2}}\;.
\label{Q1}
\end{align}
These estimations are traced back to the Meixner polynomials. The
factor $\sqrt{m^1+1}$ in (\ref{Q1}) is particularly remarkable. Any
other Taylor subtraction (e.g.\ with pre-factors $\sqrt{m^1}$ or
$\sqrt{m^1{+}2}$) would kill the renormalisation proof. 

These discrete Taylor subtractions are used in the integration from
$\Lambda_0$ down to $\Lambda$ in prescriptions like (\ref{A4}):
\begin{align}
&- \int_{\Lambda}^{\Lambda_0} 
\frac{d \Lambda'}{\Lambda'} \,
\sum_{p \in \mathbb{N}^2}  \left(~~~
\parbox{40mm}{\begin{picture}(20,15)
       \put(0,0){\epsfig{file=a24,bb=71 630 174 676}}
       \put(-4,12){\mbox{\footnotesize$m$}}
       \put(-1,6){\mbox{\footnotesize$m$}}
       \put(34,9){\mbox{\footnotesize$k$}}
       \put(36,4){\mbox{\footnotesize$k$}}
       \put(4,-1){\mbox{\footnotesize$l$}}
       \put(30,-1){\mbox{\footnotesize$l$}}
       \put(5,14){\mbox{\footnotesize$n$}}
       \put(29,15){\mbox{\footnotesize$n$}}
       \put(12,8){\mbox{\footnotesize$p$}}
       \put(22,8){\mbox{\footnotesize$p$}}
   \end{picture}}
-~~
\parbox{40mm}{\begin{picture}(20,15)
       \put(0,0){\epsfig{file=a24,bb=71 630 174 676}}
       \put(-4,12){\mbox{\footnotesize$m$}}
       \put(-1,6){\mbox{\footnotesize$m$}}
       \put(34,9){\mbox{\footnotesize$k$}}
       \put(36,4){\mbox{\footnotesize$k$}}
       \put(4,-1){\mbox{\footnotesize$l$}}
       \put(30,-1){\mbox{\footnotesize$l$}}
       \put(5,14){\mbox{\footnotesize$n$}}
       \put(29,15){\mbox{\footnotesize$n$}}
       \put(8,12){\mbox{\footnotesize$0$}}
       \put(25,12){\mbox{\footnotesize$0$}}
       \put(8,2){\mbox{\footnotesize$0$}}
       \put(25,2){\mbox{\footnotesize$0$}}
       \put(12,8){\mbox{\footnotesize$p$}}
       \put(22,8){\mbox{\footnotesize$p$}}
   \end{picture}}
\right)\![\Lambda']
\nonumber
\\*[1ex]
&\quad = \displaystyle \int_{\Lambda}^{\Lambda_0} \frac{d
  \Lambda'}{\Lambda'} 
\int_{\Lambda'}^{\Lambda_0} \frac{d
  \Lambda''}{\Lambda''}\sum_{p \in \mathbb{N}^2}
\Big( (Q_{np;pn}-Q_{0p;p0})(\Lambda') Q_{lp;pl}(\Lambda'') 
\nonumber
\\*[-2ex]
&\hspace*{12em} + Q_{0p;p0}(\Lambda')
(Q_{lp;pl}-Q_{0p;p0})(\Lambda'')
\Big) \sim \frac{C(\|n\|+\|l\|)}{\theta \Omega^2 \Lambda^2}\,.
\end{align}
The cut-off $\|n\|,\|l\|\leq 2 \theta \Lambda^2$ leads to (\ref{pc}).
\hfill $\square$\bigskip

Thus, replacing (similar as in the BPHZ subtraction) in planar $2$-
and $4$-point functions the propagators by reference propagators at
zero-indices and an irrelevant part, we have
\begin{align}
A^{(V,1,0)}_{\di{m^1}{m^2}\di{n^1}{n^2};\di{k^1}{k^2}\di{l^1}{l^2}}
  &= \Big\{A^{(V,1,0)}_{\di{0}{0}\di{0}{0};\di{0}{0}\di{0}{0}}
  + m^1 \Big(
  A^{(V,1,0)}_{\di{1}{0}\di{0}{0};\di{0}{0}\di{1}{0}}
  -A^{(V,1,0)}_{\di{0}{0}\di{0}{0};\di{0}{0}\di{0}{0}}\Big)
+ n^1 \Big(
  A^{(V,1,0)}_{\di{0}{0}\di{1}{0};\di{1}{0}\di{0}{0}}
  -A^{(V,1,0)}_{\di{0}{0}\di{0}{0};\di{0}{0}\di{0}{0}}\Big) \nonumber
  \\*
  &
+  m^2 \Big( A^{(V,1,0)}_{\di{0}{1}\di{0}{0};\di{0}{0}\di{0}{1}}
  -A^{(V,1,0)}_{\di{0}{0}\di{0}{0};\di{0}{0}\di{0}{0}}\Big)
+ n^2 \Big(
  A^{(V,1,0)}_{\di{0}{0}\di{0}{1};\di{0}{1}\di{0}{0}}
  -A^{(V,1,0)}_{\di{0}{0}\di{0}{0};\di{0}{0}\di{0}{0}}\Big)\Big\}
  \delta_{m^1l^1} \delta_{n^1k^1} \delta_{m^2l^2} \delta_{n^2k^2}
\nonumber
  \\
  &+ A^{(V,1,0)}_{\di{1}{0}\di{1}{0};\di{0}{0}\di{0}{0}}
  \big(\sqrt{k^1l^1} \delta_{m^1+1,l^1} \delta_{n^1+1,k^1}
  \delta_{m^2l^2} \delta_{n^2k^2}
+ \sqrt{m^1n^1} \delta_{m^1-1,l^1}
  \delta_{n^1-1,k^1} \delta_{m^2l^2} \delta_{n^2k^2}\big)
\nonumber
  \\*
  &
+ A^{(V,1,0)}_{\di{0}{1}\di{0}{1};\di{0}{0}\di{0}{0}}
  \big(\sqrt{k^2l^2}
\delta_{m^2+1,l^2} \delta_{n^2+1,k^2}
  \delta_{m^1l^1} \delta_{n^1k^1}
+ \sqrt{m^2n^2} \delta_{m^2-1,2^1}
  \delta_{n^2-1,k^2} \delta_{m^1l^1} \delta_{n^1k^1}\big)
\nonumber
  \\*
  & 
+ \text{ irrelevant part}\;,
\\
A^{(V,1,0)}_{m_1n_1;\dots;m_4n_4} &=
A^{(V,1,0)}_{00;\dots;00} \Big(\tfrac{1}{6} \delta_{n_1m_2}
  \delta_{n_2m_3} \delta_{n_3m_4} \delta_{n_4m_1} + 5 \text{
    perms}\Big) + \text{ irrelevant part}\;.
\end{align}
We conclude that there are four independent relevant/marginal interaction 
coefficients:  
\begin{align}
\rho_1 &=A^{(V,1,0)}_{\di{0}{0}\di{0}{0};\di{0}{0}\di{0}{0}}\;, &
\rho_2 &=A^{(V,1,0)}_{\di{1}{0}\di{0}{0};\di{0}{0}\di{1}{0}}-
A^{(V,1,0)}_{\di{0}{0}\di{0}{0};\di{0}{0}\di{0}{0}}\;, &
\rho_3 &=A^{(V,1,0)}_{\di{1}{0}\di{1}{0};\di{0}{0}\di{0}{0}}\;, &
\rho_4 &=A^{(V,1,0)}_{\di{0}{0}\di{0}{0};\di{0}{0}\di{0}{0};
\di{0}{0}\di{0}{0};\di{0}{0}\di{0}{0}}\;.
\label{rhodef}
\end{align}
At $\Lambda=\Lambda_0$ we recover the same index structure as in the
initial action (\ref{Sm}), (\ref{Gm}), identifying
$\rho_a[\Lambda_0]\equiv \rho^0_a$ as functions of the coefficients
$\mu_0,\theta,\Omega,\lambda$. This is a first indication that our
model will be renormalisable. However, we have to remove the cut-off
by sending $\Lambda_0 \to \infty$.

\section{Removal of the cut-off}
\label{removal}

For given data $\Lambda_0,\rho_a^0$, the integration of the Polchinski
equation yields the coefficients
$A^{(V,B,g)}_{m_1n_1;\dots;m_Nn_N}[\Lambda,\Lambda_0,\rho_a^0]$ and
thus, via (\ref{rhodef}), $\rho_b[\Lambda,\Lambda_0,\rho_a^0]$. Now,
according to Section~\ref{integration}, in particular (\ref{A4}), we
keep $\rho_b[\Lambda_R,\Lambda_0,\rho_a^0]$ \emph{constant} when varying
$\Lambda_0$. This leads to the identity
\begin{align}
L[\phi,\Lambda_R,\Lambda_0',\rho^0[\Lambda_0']] &-
  L[\phi,\Lambda_R,\Lambda_0'',\rho^0[\Lambda_0'']] =
  \int_{\Lambda_0''}^{\Lambda_0'} \frac{d\Lambda_0}{\Lambda_0} \;
  R[\phi,\Lambda_R,\Lambda_0,\rho^0[\Lambda_0]]\;,
\label{LR0}
\\
R[\phi,\Lambda,\Lambda_0,\rho^0] &:= \Lambda_0 \frac{\partial
    L[\phi,\Lambda,\Lambda_0,\rho^0]}{ \partial \Lambda_0} 
- \sum_{b=1}^4 
H^b[\Lambda,\Lambda_0,\rho^0]\;
\Lambda_0 \frac{\partial
    \rho_b[\Lambda,\Lambda_0,\rho^0]}{ \partial \Lambda_0}\;,
\label{Rphi}
\\
H^b[\Lambda,\Lambda_0,\rho^0]
&:= \sum_{a=1}^4 \frac{\partial L[\phi,\Lambda,\Lambda_0,\rho^0]}{ \partial
    \rho_a^0} \frac{\partial \rho_a^0}{\partial
    \rho_b[\Lambda,\Lambda_0,\rho^0]} \;.
\end{align}
From (\ref{polch}) one derives Polchinski-like differential 
equations for the coefficients of $R$ and $H^a$:
\begin{align}
\Lambda \frac{\partial R}{\partial \Lambda} &= M[L,R] {-} \sum_{a=1}^4
H^a M_a[L,R]\;, &
\Lambda \frac{\partial H^a}{\partial \Lambda} 
&= M[L, H^a ] {-}  \sum_{b=1}^4 H^b M_b[L, H^a]\;,
\label{M}
\end{align}
for certain functions $M, M_a$ which are linear in the second
argument. We only have initial conditions at $\Lambda_0$ for these
coefficients, thus the integration must always be performed from
$\Lambda_0$ down to $\Lambda$. Fortunately, there are (by
construction) remarkable cancellations in the rhs of (\ref{M}) so that
relevant contributions never appear. One proves
\begin{prp}
\begin{align}
\big|H^{a(V,B,g)}_{
    m_1n_1;\dots;m_Nn_N}[\Lambda,\Lambda_0,\rho_0] \big| 
\leq \big(\sqrt{\theta} \Lambda\big)^{(4-N-2\delta^{a1})
    +4(1-B-2g)}
P^{2V+1+\delta^{a4} -\frac{N}{2}}\Big[\ln
  \frac{\Lambda_0}{\Lambda_R}\Big]\;,
\\
\big|R^{(V,B,g)}_{
    m_1n_1;\dots;m_Nn_N}[\Lambda,\Lambda_0,\rho_0] \big|
\leq
\Big(\frac{\Lambda^2}{\Lambda_0^2}\Big)
\big(\sqrt{\theta} \Lambda\big)^{(4-N)+4(1-B-2g)}
\,P^{2V-\frac{N}{2}}\Big[\ln\frac{\Lambda_0}{\Lambda_R}\Big]\;.
\label{Rest}
  \end{align}
\end{prp}
We give the main ideas of the proof of (\ref{Rest}). First,
$R^{(1,1,0)}_{m_1n_1;\dots;m_4n_4} \equiv 0$, because the $\phi^4$-vertex is
scale-independent, which leads to a vanishing coefficient according to
(\ref{Rphi}). Then, as $R^{(1,1,0)}_{m_1n_1;\dots;m_4n_4}$ appears in
each term on the rhs of the first differential equation (\ref{M}) for
the 2-vertex six-point function and the 1-vertex two-point function,
the coefficients 
$R^{(2,1,0)}_{m_1n_1;\dots;m_6n_6}, 
R^{(1,1,0)}_{m_1n_1;m_2n_2}$ and  
$R^{(1,2,0)}_{m_1n_1;m_2n_2}$ are $\Lambda$-independent.
Next, one derives e.g.\ 
\begin{align}
R^{(2,1,0)}_{m_1n_1;\dots;m_6n_6}[\Lambda_0,\Lambda_0,\rho^0]= -
\Big(\Lambda \frac{\partial}{\partial \Lambda} 
A^{(2,1,0)}_{m_1n_1;\dots;m_6n_6}[\Lambda,\Lambda_0,\rho^0]
\Big)_{\Lambda=\Lambda_0} \sim \frac{C}{\theta \Lambda_0^2}\,,
\label{R6}
\end{align}
where the scaling behaviour follows from (\ref{pc}). Since the first
differential equation (\ref{M}) is linear in $R$ and relevant
coefficients are projected away, the relative factor
$\frac{\Lambda^2}{\Lambda_0^2}$ between $|A[\Lambda]|$ and
$|R[\Lambda]|$ which first appears in (\ref{R6}) and similarly in
$R^{(1,1,0)}_{m_1n_1;m_2n_2}, R^{(1,2,0)}_{m_1n_1;m_2n_2}$ survives to
all $R$-coefficients. By integration of (\ref{LR0}) we thus obtain
\begin{thm}
The duality-covariant noncommutative $\phi^4$-model
is (order by order in the coupling constant) renormalisable 
\\[1ex]
\begin{tabular}{lp{0.93\textwidth}}
-- & by an
adjustment of the initial coefficients $\rho^0_a[\Lambda_0]$ to give
renormalised constant couplings
$\rho_a^R=\rho_a[\Lambda_R,\Lambda_0,\rho_b^0[\Lambda_0]]$, and 
\\[1ex]
-- &by 
the corresponding integration of the flow equations.
\end{tabular}
\\[2ex]
The limit
  $\displaystyle A^{(V,B,g)}_{m_1n_1;\dots;m_Nn_N}[\Lambda_R,\infty]
  :=\lim_{\Lambda_0 \to \infty}
  A^{(V,B,g)}_{m_1n_1;\dots;m_Nn_N}
  [\Lambda_R,\Lambda_0,\rho^0[\Lambda_0]]$ of the
  expansion coefficients of the effective action
  $L[\phi,\Lambda_R,\Lambda_0,\rho^0[\Lambda_0]]$ exists and satisfies
  \begin{align}
\Big| (2\pi\theta)^{\frac{N}{2}-2}
  A^{(V,B,g)}_{m_1n_1;\dots;m_Nn_N}[\Lambda_R,\infty] &-
  (2\pi\theta)^{\frac{N}{2}-2}
  A^{(V,B,g)}_{m_1n_1;\dots;m_Nn_N}
[\Lambda_R,\Lambda_0,\rho^0]  \Big|
\nonumber
\\
& \leq \frac{\Lambda_R^{6-N}}{\Lambda_0^2}
\Big(\frac{1}{\theta^2\Lambda_R^4}\Big)^{B+2g-1}
  \,P^{2V-\frac{N}{2}}
\Big[\ln \frac{\Lambda_0}{\Lambda_R}\Big]\;.
\end{align}
\end{thm}

\section{ Renormalisation group equation} 
\label{RGE}

Knowing the relevant/marginal couplings, we can compute Feynman graphs
with sharp matrix cut-off $\mathcal{N}$. The most important question
concerns the $\beta$-function appearing in the renormalisation group
equation which describes the cut-off dependence of the expansion
coefficients $\Gamma_{m_1n_1;\dots;m_Nn_N}$ of the effective action
when imposing normalisation conditions for the relevant and marginal
couplings. We have \cite{Grosse:2004by}
\begin{align}
\lim_{\mathcal{N}\to \infty} 
\Big(\mathcal{N} \frac{\partial}{\partial \mathcal{N}} 
+ N \gamma + \mu_0^2 \beta_{\mu_0} \frac{\partial}{\partial \mu_0^2} 
+ \beta_\lambda \frac{\partial}{\partial \lambda} 
+ \beta_\Omega \frac{\partial}{\partial \Omega} \Big) 
\Gamma_{m_1n_1;\dots;m_Nn_N}[\mu_0,\lambda,\Omega,\mathcal{N}] = 0\;,
\end{align}
where
\begin{align}
\beta_\lambda &=  \mathcal{N} 
\frac{\partial }{\partial \mathcal{N}}   
\Big( \lambda[\mu_{\text{phys}},\lambda_{\text{phys}},
\Omega_{\text{phys}},\mathcal{N} ]\Big) \;, &
\beta_\Omega &=  
 \mathcal{N} \frac{\partial }{\partial \mathcal{N}} 
\Big( \Omega[\mu_{\text{phys}},\lambda_{\text{phys}},
\Omega_{\text{phys}},\mathcal{N} ]\Big) \;,
\nonumber
\\
\beta_{\mu_0} &=  \frac{\mathcal{N} }{\mu_0^2}  
\frac{\partial }{\partial \mathcal{N}} 
\Big( \mu_0^2[\mu_{\text{phys}},\lambda_{\text{phys}},
\Omega_{\text{phys}},\mathcal{N} ]\Big) \;, &
\gamma &= \mathcal{N} \frac{\partial }{\partial \mathcal{N}}
\Big( \ln \mathcal{Z}[\mu_{\text{phys}},\lambda_{\text{phys}},
\Omega_{\text{phys}},\mathcal{N} ]\Big) \;.
\end{align}
Here, $\mathcal{Z}$ is the wavefunction renormalisation. To one-loop
order we find \cite{Grosse:2004by}
\begin{align}
\beta_\lambda &=  \frac{\lambda_{\text{phys}}^2}{48 \pi^2} 
\frac{(1{-}\Omega_{\text{phys}}^2)}{(1{+}\Omega_{\text{phys}}^2)^3} \;,
&
\beta_\Omega &=  
\frac{\lambda_{\text{phys}} \Omega_{\text{phys}}}{96 \pi^2} 
\frac{(1{-}\Omega_{\text{phys}}^2)}{(1{+}\Omega_{\text{phys}}^2)^3}\;,
\nonumber
\\
\beta_{\mu_0} &= - \dfrac{\lambda_{\text{phys}}
\Big(4\mathcal{N}\ln(2)
+ \frac{(8{+} \theta\mu_{\text{phys}}^2)\Omega^2_{\text{phys}}}{
(1{+}\Omega_{\text{phys}}^2)^2} \Big)}{48 \pi^2 \theta
  \mu_{\text{phys}}^2 (1{+}\Omega_{\text{phys}}^2) }\;,
&
\gamma &=  \frac{\lambda_{\text{phys}} }{96 \pi^2} 
\frac{\Omega^2_{\text{phys}}}{(1{+}\Omega_{\text{phys}}^2)^3}\;.
\end{align}
There are two remarkable special cases. First, for $\Omega=1$, which
corresponds to a self-dual model according to (\ref{Sdual}), we have 
$\beta_\lambda = \beta_\Omega= 0$. This is true to all orders for 
$\beta_\Omega$ and conjectured for $\beta_\lambda$ due to the
resemblance of the duality-invariant theory with the exactly
solvable models discussed in \cite{Langmann:2003if}.
Second, $\beta_\Omega$ also vanishes in the limit $\Omega\to 0$, which
defines the standard noncommutative $\phi^4$-quantum field theory.
Thus, the limit $\Omega \to 0$ exists at least at the one-loop level.

\end{document}